\documentclass[sigconf]{acmart}

\usepackage{graphicx}
\graphicspath{{figures/}}
\usepackage{multirow}
\usepackage{balance}

\AtBeginDocument{%
  }

\newcommand{\tool}{Proof Blocks}

\newcommand{\pl}{PrairieLearn}

\begin{document}

\title{Efficiency of Learning from Proof Blocks Versus Writing Proofs}

%\author{Seth Poulsen, Yael Gertner, Benjamin Cosman, Matthew West, Geoffrey L.
%Herman}
%\email{{sethp3, ygertner, bcosman, mwest, glherman}@illinois.edu}
%%\orcid{1234-5678-9012}
%\affiliation{%
%	\institution{University of Illinois Urbana-Champaign}
%	%  \streetaddress{P.O. Box 1212}
%	%  \city{Dublin}
%	%  \state{Ohio}
%	%  \postcode{43017-6221}
%	\country{}
%}
%
\author{Seth Poulsen}
%\author{Yael Gertner}
%\author{Benjamin Cosman}
%\author{Geoffrey L. Herman}
%\author{Matthew West}
\email{sethp3@illinois.edu}
%\email{ygertner@illinois.edu}
\affiliation{%
	\institution{University of Illinois Urbana-Champaign}
	%  \streetaddress{P.O. Box 1212}
	\country{}
}

\author{Yael Gertner}
\email{ygertner@illinois.edu}
\affiliation{%
	\institution{University of Illinois Urbana-Champaign}
	%  \streetaddress{P.O. Box 1212}
	\country{}
}

\author{Benjamin Cosman}
\email{bcosman@illinois.edu}
\affiliation{%
	\institution{University of Illinois Urbana-Champaign}
	%  \streetaddress{P.O. Box 1212}
	\country{}
}

\author{Matthew West}
\email{mwest@illinois.edu}
\affiliation{%
	\institution{University of Illinois Urbana-Champaign}
	%  \streetaddress{P.O. Box 1212}
	\country{}
}

\author{Geoffrey L. Herman}
\email{glherman@illinois.edu}
\affiliation{%
	\institution{University of Illinois Urbana-Champaign}
	%  \streetaddress{P.O. Box 1212}
	\country{}
}
%
%\author{Anon}
%\email{anon@anon.edu}
%%\orcid{1234-5678-9012}
%\affiliation{%
%	\institution{University of Anonymous}
%	%  \streetaddress{P.O. Box 1212}
%	%  \city{Dublin}
%	%  \state{Ohio}
%	%  \postcode{43017-6221}
%	\country{Anonymous Country}
%}

%\renewcommand{\shortauthors}{Poulsen et al.}
\renewcommand{\shortauthors}{Seth Poulsen, Yael Gertner, Benjamin Cosman, Matthew
West, \& Geoffrey L. Herman}

\begin{abstract}
\tool{} is a software tool that provides students with a scaffolded
proof-writing experience, allowing them to drag and drop prewritten proof lines
into
the correct order instead of starting from scratch. In this paper we
describe a randomized controlled trial designed to
measure the learning gains of using \tool{} for students learning proof by
induction. The study participants were 332
students recruited after completing the first month of their discrete
mathematics course.
Students in the study took a pretest and read lecture notes on proof by induction,
completed a brief
(less than 1 hour) learning activity, and then returned one week later to
complete the posttest.
Depending on the experimental condition that each student was assigned to, they
either completed only \tool{} problems, completed some \tool{} problems and
some written proofs, or completed only written proofs for their learning
activity.
We find that students in the early phases of learning about proof by induction
are able to learn just as much from reading lecture notes and using \tool{} as by
reading lecture notes and writing proofs from
scratch, but in far less time on task. This finding complements previous findings
that \tool{} are useful
exam questions and are viewed positively by
students.
\end{abstract}




\maketitle

\section{Introduction and Background}
\begin{figure}
\includegraphics[width=\columnwidth]{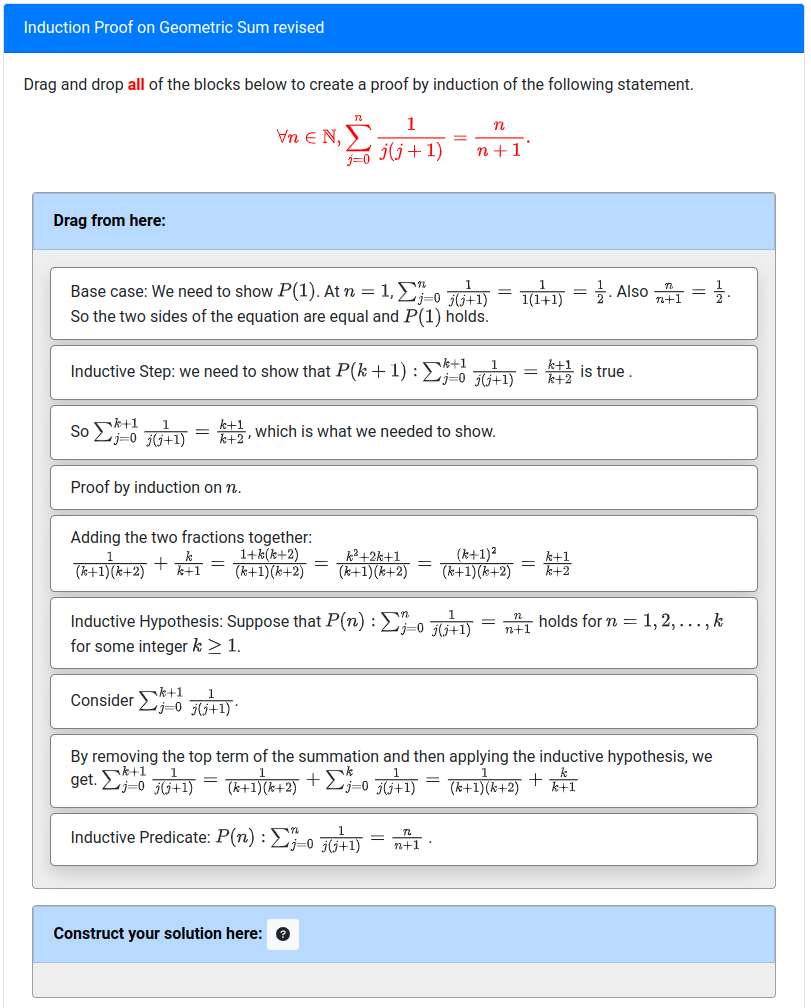}
\caption{Example \tool{} problem from the Proof Blocks learning activity.
Students are provided with the lines of the proof in scrambled order,
and they must drag and drop them back into the correct order. They are
able to receive instant feedback on their submission.}
\label{fig:example}
\vskip -2em
\end{figure}
There is little research on teaching interventions for
writing mathematical proofs, despite it being one of the most difficult and
important aspects of computing theory to
learn~\cite{goldman2008identifying,stylianides2017research}. Even when
students have all the content knowledge needed to construct a proof, they still
struggle to write proofs, showing that students need scaffolding to
help them through the
process~\cite{weber2001student,vygotsky1978mind}. To address this need, we've
created \tool, a software tool that allows students to drag and drop
prewritten proof lines into the correct order rather than having
to write a proof completely on their own (see Figure~\ref{fig:example} for an
example \tool{} problem). In
this paper, we detail a study designed to measure the learning gains of
students practicing their proof-writing skills using \tool.
%An example \tool{} problem can be seen in Figure~\ref{fig:example}.
%To create a
%\tool{} problem, an instructor writes the lines of the proof and specifies
%which lines of the proof logically depend on which other lines of the
%proof, thus forming a directed acyclic graph (DAG)~\cite{poulsen2022proof}. The
%lines are randomly rearranged, and students must
%drag and drop the lines back into the correct order. Each time a student
%submits an answer, they can receive immediate feedback on where their
%submission went wrong.
While \tool{} have been shown to be useful in
assessment~\cite{poulsen2021evaluating}, there has been no previous research on
their utility for student learning. We
hypothesize that \tool{} problems can also improve learning, because the
prewritten lines provide a scaffold that help students focus on the sub-task
of constructing logical arguments. To test this hypothesis, we performed a
randomized controlled trial to compare the learning gains of
students who completed a \tool{} learning activity to students who completed a
proof writing learning activity, and students who completed a learning activity
with some proof writing and some \tool{} problems. The authors' prior
work reviews other
software tools providing visual methods of constructing
proofs~\cite{poulsen2022proof}.

\vspace{-1em}
\subsection{Research on Teaching and Learning Proofs}
Many threads of research seek to illuminate students' understandings
and misunderstandings about proofs
\cite{selden2008overcoming,stylianides2016proof,stylianides2017research}.
One thread establishes that, as they learn, students go through different
phases in how they are able to think about solving proof
problems~\cite{weber2004semantic}.
Another study demonstrated that even when students had all of the knowledge
required to write a proof and were able to apply that knowledge in other types
of questions, they were still unable to write a proof~\cite{weber2001student},
thus highlighting that performing all aspects of proof writing at once is a
separate challenge than learning the component skills of proof writing.

On the other hand, there is little research on concrete educational
interventions for improving the proof-learning
process~\cite{hodds2014self,stylianides2017research}.
Based on their review of the literature on teaching and learning proofs,
Sylianides and Stylianides concluded that ``more intervention-oriented studies
in the area of proof are sorely needed ~\cite{stylianides2017research}.''

%There have been a few other studies which taken a non-experimental approach to
%proof education interventions. These
%studies give a rich reflection on the development of novel instructional
%practices, and qualitative data on how they were received by the students.
%Harel~\cite{harel2001development} makes argument that traditionally, proof by
%induction has focused too much on the result of recursive calculations, and not
%enough about the structural patterns that occur when applying recursive rules.
%Larsen and Zandieh~\cite{larsen2008proofs} create excercises that guide
%students
%through the process of mathematical discovery in a way that mimics how it was
%originally discovered (guided reinvention) and share details of an account of
%using this approach with students.

There have been a few experimental studies seeking to improve student
understanding of proofs~\cite{weber2012generic,malek2011effect,roy2014evaluating}.
Hodds et al.~\cite{hodds2014self} showed that
training students to engage more with proofs by using self-explanation methods
increased student comprehension of proofs in a lasting way.
%\textcolor{blue}{not sure the following sentence is needed or is something we
%can defend rigorously. Consider cutting}\tool{} problems similarly force
%deliberate engagement with proof content, as close reading is necessary to
%determine the correct arrangement of lines.
There have also been interventions that focus on helping students understand the
need for proofs, as students often believe empirical arguments without seeing
the need for
proof~\cite{stylianides2017research,jahnke2013understanding,brown2014skepticism,
stylianides2009facilitating}. A few studies have taken a
non-experimental approach, giving reflections on the development of novel
instructional practices~\cite{harel2001development,larsen2008proofs}.
%Though there have been experimental studies on aiding proof understanding, and
%non-experimental studies on helping students learn to write proofs,
To our knowledge, our study is the first \emph{experimental} study
%(as opposed to non-experimental)
on an
intervention to improve student's abilities to \emph{write} (as opposed to
understand) mathematical proofs.

\vspace{-1em}
\subsection{Parsons Problems}
\tool{} were inspired by Parsons problems, which similarly scaffold students'
learning of programming by scrambling prewritten lines of programs
\cite{parsons2006parson}. Parsons problems have been studied for their
desirable properties both in assessment and
learning~\cite{denny2008evaluating,du2020review,ericson2017solving,
ericson2018evaluating,weinman2021improving}.
Denny et al.~\cite{denny2008evaluating} showed that Parsons problems are
easier to
grade than free-form code writing questions, and yet still offer rich
information about student knowledge. The same is true of
\tool{} problems relative to free-form proof-writing
questions~\cite{poulsen2021evaluating}.

Ericson et al.
\cite{ericson2017solving,ericson2018evaluating}
performed a series of randomized controlled experiments to compare the learning
gains of students using Parsons problems to the learning gains of students
doing other learning activities.
First, they compared the learning gains of students using Parsons problems to
students fixing code and writing code from scratch. In a posttest measurement
of both Parsons problems and code writing questions, they found that students
learned equal amounts
across all conditions. In their next experiment, they compared adaptive Parsons
problems, Parsons problems, code writing, and a control condition where
students solved Parsons problems on a topic unrelated to the posttest topic.
They found that students who completed the Parsons problems and adaptive
Parson's problems learned the most.
Work is in progress  to replicate this work across many
universities~\cite{ericson2022planning}.
Similar to the work of Ericson et al, the study which we report on is designed
to confirm if \tool{} help accelerate the learning process of writing proofs.

%Many variants of Parsons problems have appeared in classrooms and in the
%research literature. The \tool{} problems used in our study most closely
%resemble Parsons problems with relative line-based feedback and no
%distractors~\cite{du2020review}.

\vspace{-1em}
\subsection{Research Questions}
%In courses that teach mathematical proofs, students often learn to write
%mathematical proofs by reading proofs from a book, watching their instructors
%or teaching assistants writing mathematical proofs, and then attempting to
%write mathematical proofs on their own \textcolor{blue}{citation? evidence?}.
One traditional proof learning activity is
for students to attempt to write mathematical
proofs on their own and then view an example solution to assess
their progress. We call this a \emph{proof-writing} activity. Due to its
widespread use, we feel that a proof-writing activity makes the most
appropriate baseline for comparing the learning gains of \tool{}.
Because \tool{} are a scaffolded version of proof writing, we felt that
completing some \tool{} problems and then some proof-writing problems would be
an effective learning activity. We call this a \emph{hybrid}
learning activity. In this study, we seek to quantify the difference in
learning gains between students who complete a learning activity containing
only \tool{} problems, a learning activity containing only proof-writing
problems, and students who complete a hybrid learning activity.

We go in to this study with the following research questions:

\begin{quote}
\textbf{RQ1}: What are the differences in learning gains between students
who complete a \tool{} activity, proof-writing activity, or hybrid activity? \\
%\textbf{RQ2}: How does time on task vary between the experimental conditions?
\textbf{RQ2}: How does time on task vary between learning conditions, and how
does it relate to learning gains?
\end{quote}

We originally hypothesized that the learning gains for students in the hybrid
activity would be the most time efficient, meaning that students in the hybrid
activity would learn the most relative to the amount of time spent on task.

We answer these questions in the context of a proof by induction learning
activity. Proof by induction is a perfect topic because it has been rated as
one of the most important and difficult topics in a discrete math course
~\cite{goldman2008identifying}. Proof by induction is
so important because it is a prevalent technique that is referred to in upper
level courses and it employs recursive thinking as well as
precise logical arguments. Therefore, helping students gain mastery of this
topic is impactful not only for their success in the course but also for their
later CS courses.
%Students struggle with this topic and so figuring out what kind of help they
%need is helpful. It is also an important topic so it is important to make sure
%students understand it. It is also a topic that is taught with a very clear
%outline for grading and has similar problems that are easy to practice on
%and then generalize to the one at test.
Prior work shows that students lack key conceptual knowledge related to
induction~\cite{harel2001development,brown2008exploring,ron2004use}.
%Students often think they understand while reading the proofs but then cannot
%reproduce a proof for a new problem~\cite{weber2001student}.

\vspace{-1em}
\section{Experimental Design}
We used a between-subjects experimental design.
To control for confounding variables, we ran
our study as a controlled lab study rather than as part of a course.
We recruited students from the Discrete Mathematics course in our department
who already had knowledge of some kinds of proofs and used \tool{} to teach
them a new kind of
proof.
All students who participated took a pretest, read through some lecture notes
about proof by
induction, completed a learning activity (proof-writing, \tool, or hybrid), and
then took a practice test (see Table~\ref{tab:studydesign}).
As soon as a student finished one section, they were allowed to proceed to the
next.  We also
invited them all to participate in a posttest one week later. The
pretest, practice test, and posttests were exactly identical, all containing
the exact same two proof-writing problems:
proving the correctness of closed-form formula for a finite
sum, and proving the correctness of a closed-form solution to a recursively
defined function. We expected the second topic to be more difficult.
While the students were shown the
example solutions for all exercises they complete during the learning
activities, they were not shown the example solutions for any of the test
questions.

\begin{table}
	\centering
	\begin{tabular}{c|c|c}
		Group A & Group B & Group C \\ \hline \hline
		Pretest & Pretest & Pretest \\
		Lecture Notes & Lecture Notes & Lecture Notes
		\\
		\tool{} & Hybrid Activity &	Proof Writing \\
		Practice Test & Practice Test & Practice Test \\
		\hline
		\multicolumn{3}{c}{One Week}  \\
		\hline
		Posttest & Posttest & Posttest
	\end{tabular}
	\caption{Design of the learning experiment. Students were free to move on
	as soon as they finish a particular portion of the activity}
	\label{tab:studydesign}
\vspace{-4em}
\end{table}

\vspace{-1em}
\subsection{Experiment Environment}
All students completed their learning activities and tests in \\\pl,
%an Open source Homework and Exam Platform, which we will call \pl{} for
%anonymization purposes.
a problem-driven online learning system~\cite{prairielearn}.
Since \pl{} automatically keeps track of
when a user opens and closes each assessment and when they submit an answer to
each problem, we were able to track and analyze the
amount of time that students spend on each portion of the study.
Since Fall 2021, all sections of Discrete Mathematics have submitted their
homework and completed their exams through \pl{}, so the students we recruited
for the study are already familiar with the platform and its user
interface.
Students wrote their proofs in a text entry box
that supported markdown and LaTeX, but were told that using plain text (for
example, spelling out ``sum from i=0 to n" instead of using $\sum_{i=0}^n$) was
acceptable, as they were
not expected to learn LaTeX for the course.
To control the student learning environment for our study, we used our
University's computer based testing facility, which provides a locked down
environment where student could only access the
assessment that they were working on~\cite{zilles2019every}. Students could
choose to
complete the learning activity at any
point over the period of a few days.

%To control the student learning environment for our study, we used the
%University's Computer Based Testing Facility
%(CBTF)~\cite{zilles2019every}, which provides a proctored and locked down
%environment where student work on lab computers where they can only access the
%assessment which they are supposed to be working on. The CBTF allows flexible
%scheduling so that students can choose to complete the learning activity at any
%point over the period of a few days.

\subsection{Experimental Subjects}
Students in the Discrete Mathematics course in our department learn proof by
induction sometime in the middle of the semester. Thus, we could recruit
students for our study roughly a month into the semester, after they have
learned the basics of writing proofs, but before their course has covered
proof by induction. Since all parts of the experiment were complete before the
students covered proof by induction in class, the students had little
to no motivation to study the material outside of the context of the study,
helping with the validity of the experiment. The Discrete Mathematics course in
our department is typically taken by first year students in the computer
science and computer engineering majors or computer science minor.
Introductory programming and calculus are prerequisites.

We ran a pilot study during Fall 2021 with 5 students to test our materials,
and students were given a gift card as compensation for their participation.
For the main study during the Spring of 2022, we offered students one homework
assignment of extra credit for participating in each day of the study (one for
the learning activity, another for the posttest). Due to
constraints of \pl, it was much easier to pre-assign all
eligible students to an experimental condition before they elected to
participate. This resulted in a small variation in the population of sizes for
each treatment. We had 451
students participate in the learning activity. Of these, 353 students showed up
the following week to complete the posttest. As allowed by the terms of our
research protocol, 13 subject in this pool opted to not have their data used
for purposes of the research project. Another 8 participants did not complete
all of the questions. After removing these, we were left with a final data
set of 332 students. Broken down by experimental condition, 107 of 138 (77.5\%)
of students who started in the Proof Blocks condition, 112 of 160 (70\%) of
students who started in the Hybrid learning condition, and 113 of 153 (73.8\%)
of students who started in the Written Proofs condition were included in the
final data set.

\vspace{-1em}
\subsection{Learning Activity Materials}
We designed the experiment to maximize the similarity between learning
activities, so that the only difference between the learning activities was the
way students were constructing their proofs. The
theorems that the students are proving are the exact same between the proof
writing and \tool{} problems, and the example solutions that are shown to the
students writing proofs from scratch are identical to
the proofs in the \tool{} problems. Thus, all students are provided with
all of the exact same information, the only difference is the way they
are interacting with that information.

As an extra check for the comparability of experimental groups, at the
beginning of the learning activity students were asked a single
question to gauge their level of familiarity with proof by induction: ``What
was your level of familiarity with proof by induction before today?'' with
answer choices (a) I was very familiar with proof by induction, (b) I was
somewhat familiar with proof by induction, and (c) I had never heard of proof
by induction.

Each of the learning activities consists of five problems: three
that are similar proofs to the first proof problem from the test, and two
that are similar to the second problem. The students completing the
\tool{} learning activity completed all five of these problems as \tool, while
the students in written proofs activity completed all five of these problems as
written proofs. In the hybrid activity,
students are given two \tool{} problems, then one written proof, followed by
another \tool{} problem and then another written proof.

We wanted to encourage students working on written proofs to make a good faith
attempt at writing the proof instead of just clicking through the prompt. To do
so, in
addition to the text entry box for them to write their proof in,
we added another text entry box to the problem with a prompt
for students to compare their proof to the example proof shown them after their
initial submission. The instruction
to ``compare'' is intentionally vague. We wanted to encourage the
students to make a good faith attempt at each problem, but we
did not want to give students extra scaffolding for meta-cognition above and
beyond what they were given in the \tool{} problems, to ensure that we are
fairly comparing between the learning conditions.

Students working on \tool{} were given instant feedback on their work, including
which line of their proof was the first incorrect line. This type of feedback is
commonly used in
Parsons problems, and it has
been called relative line-based feedback~\cite{du2020review}.
For more details
of the \tool{} autograder and feedback system,
see~\cite{poulsen2022proof,poulsen2022efficient}.
Students were given three tries to complete each \tool{} before being shown the
example solution.  While \tool{} does support the use of
distractors, and many instructors use them this way in the classroom, we
decided to use no distractors in this first study for simplicity, but we plan to
do so in future work.

\vspace{-1em}
\section{Data Analysis}
%In the study by Ericson et al.~\cite{ericson2017solving}, grading student work
%was straightforward as the were able to assign scores to students
%simply by running their code against a suite of unit tests. In our case,
%however the situation is a bit more complicated, because the measure on which
%we want to compare
%students is their ability to write mathematical proofs from scratch. There is
%no
%reliable way to automatically grade student proofs.

\subsection{Rubric}
There are no existing
validated rubrics for assigning student grades for mathematical proofs. Thus, as
part of our study, we designed and validated a rubric for grading student
proofs, in addition to grading all student work by hand.
While there has been some work to understand how mathematicians typically
grade proofs~\cite{miller2018mathematicians}, we have no knowledge of research
attempting to find a standard rubric.

This use of a rubric can be considered a data transformation that converts our
qualitative data (students' written proofs) to quantitative data (a numerical
score). A discussion of the methods and validity
of such transformations, and more examples, can be seen in standard books on
mixed methods research: ~\cite{creamer2017introduction,creswell2007designing}.
We started out with a rubric that had been
used for proof by induction problems in recent semesters in our department, then
edited it to remove as much human judgment as possible in the interest of
creating a reliable measure.
We then graded the student proofs
collected during our pilot study.
To start, three members of the research team independently graded two proofs
from one student. We met together with a fourth member of the
research team to come to an agreement about which rubric points should be
applied to each proof, revising our rubric as necessary.
Then three members of the research team graded proofs by more
students, and we met together to attempt to agree on the meaning of
the rubric points and clarify their wording as much as possible.
After the final round of grading the pilot study data,
we used Krippendorff's alpha to calculate an inter-rater reliability of $0.82$
over $n=56$ rubric points ($8$ proofs).
Since this was above the generally accepted threshold of
0.8~\cite{krippendorff2004reliability}, we decided that our rubric was reliable
enough to have only one member of the research team grade each student proof
moving forward. The final rubric is shown in Table~\ref{tab:codebook}.
\begin{table}
\centering
\begin{tabular}{p{2cm}|p{5.5cm}}
Proof Section & Proof Detail \\
\hline
\hline
Base Case(s)
& (1) Identify Base Case(s) \\
& (2) Prove Base Case(s)	\\
\hline
Inductive
& (3) Hypothesis is stated	\\
Hypothesis
& (4) Hypothesis is given some bound	\\
\hline
Inductive Step
& (5) Goal is Stated	\\
& (6) Expression of Size $k+1$ is decomposed into expression of size
$k$	\\
& (7) Inductive Hypothesis is applied	\\
\end{tabular}
\caption{Rubric for grading written proofs. Each detail on the rubric is
assigned the following points: 0 for not
present, 1 for partially correct, or 2 for correct. We validated our rubric by
having multiple authors grade the same proofs and iteratively refining it,
achieving a Krippendorff's alpha of $0.82$.}
\label{tab:codebook}
\vskip -3em
\end{table}

Each point on the rubric is assigned 0 for not
present, 1 for partially correct, or 2 for correct (2 questions $\times$ 7
rubric items $\times$ 2 points each  $=$ 28 points possible on the test), and
then the test score as a whole is converted to be out of 100 for ease of
reporting.
As a further measure to ensure no bias entered the grading process, when
grading the data from the main study the graders were blinded to whether
the proof they were grading came from a pretest or posttest and were blinded
as to what experimental condition the subject they were grading was under.
Of the $332 \times 4 = 1,328$ proofs in the main study data set, $1,170$ were
graded by author the first author, $49$ were graded by the second author, and
$109$ were graded by the third author.
The proof by induction familiarity survey question was also converted to a
numerical score, with `very familiar' being assigned 2, `somewhat familiar'
being assigned 1, and `never heard of' being assigned 0.

\subsection{Comparability of Experimental Groups}
Because students were randomly assigned to experimental conditions, we had
strong reason to believe a priori that the populations of
students in each experimental group were comparable, but we still ran
statistical checks to be certain.
First, we compared the student responses to the survey on the familiarity with
proof by induction. A Shapiro-Wilk test showed the data to be non-normal
($p<.001$ for all three experimental groups), so we used a Kruskal-Wallis test
to confirm that the familiarity level is similar across groups.
We fail to reject the null hypothesis that the distribution
of familiarity scores between groups are the same ($\chi^2=2.34$, $p=0.31$).
Details of the familiarity survey resultes can be seen in
Table~\ref{tab:familiarity}. We also compare the groups' pretest scores to one
another in a similar manner. A
Shapiro-Wilk test also showed that the pretest scores were non-normal ($p<.001$
for all three experimental groups), so we again use a Kruskal-Wallis test
for this comparison. We fail to reject the null hypothesis that the
distribution
of pretest scores between groups are the same ($\chi^2=0.97$, $p=0.62$).
\begin{table}
\begin{tabular}{r|c|c|c}
	Level of Familiarity& \multicolumn{3}{c}{Experimental Condition} \\
	\cline{2-4}
	 with Induction & Proof Blocks & Hybrid & Written Proofs \\ \hline
	Very Familiar  &   9    &      6     &        7 \\
	Somewhat Familiar &   37    &      30    &        29 \\
	Never Heard &   66  &        71   &         77
\end{tabular}
\caption{Breakdown of prior students knowledge by experimental group. This
	data, along with the pretest scores, confirms that each of the experimental
	groups started out roughly equal in knowledge of proof by induction.}
\label{tab:familiarity}
\vskip -2em
\end{table}
After verifying the comparability of the experimental
groups had similar baseline knowledge, we proceed with the
analysis that directly answers the research questions.

\subsection{Learning Gains}
\begin{figure*}
\includegraphics[width=\textwidth]{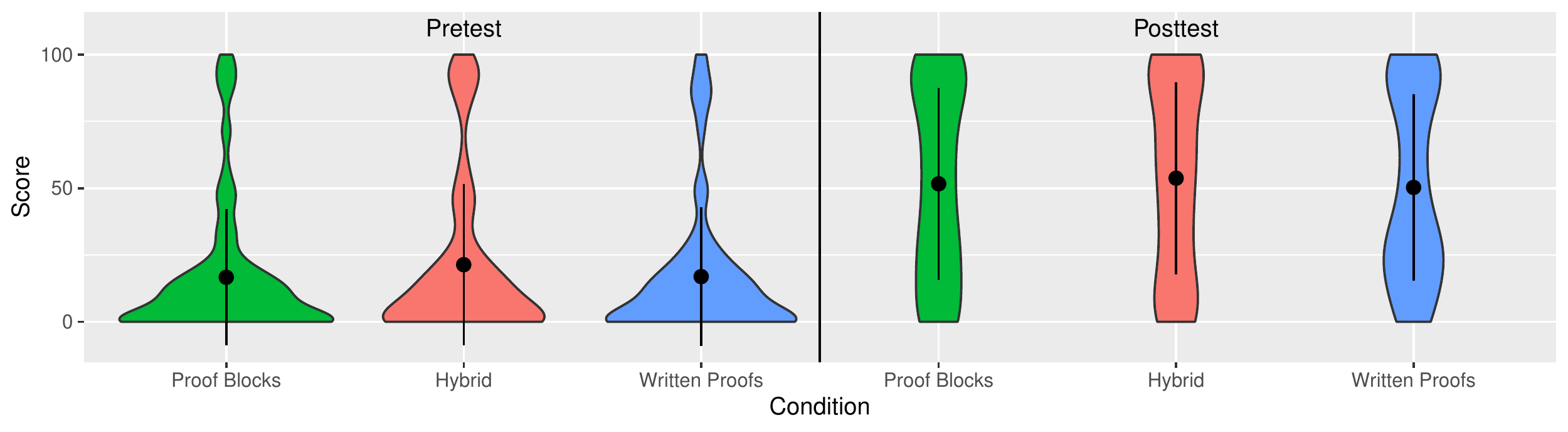}
\caption{Comparison of pretest/posttest performance across conditions. Students
in all 3 experimental groups increased their scores by between 30 and 40
percent between the pretest and the posttest. See Table~\ref{tab:preposttests}
for full score details}
\label{fig:all-violin}
\end{figure*}

To measure learning gains and answer \textbf{RQ1}, we compare the pretest
scores of each group to the posttest scores of the same group. These score
distributions can be seen in Figure~\ref{fig:all-violin}, with summary
statistics in Table~\ref{tab:preposttests}.
\begin{table*}
\begin{tabular}{c|ccccc}
& Pretest (20\%, 80\%)
& Posttest (20\%, 80\%)
& Score Increase
& Wilcox's $Q$
& Activity Time, minutes (s.d.)
\\ \hline
Proof Blocks ($n=107$) & 16.7 (0, 21.4) & 51.6 (14.3, 92.9) & 35.0 & 0.72 &
11.1 (6.3) \\
Hybrid ($n=112$) & 21.4 (0, 42.1) & 53.7 (11.4, 92.9) & 32.4 & 0.76 & 32.1
(16.0) \\
Written ($n=113$) & 16.9 (0, 21.4) & 50.3 (14.3, 92.9) & 33.4 & 0.66 & 43.4
(20.6) \\
\end{tabular}
\caption{Mean, 20\% and 80\% quantiles for the pretest and posttest scores (out
of 100), score improvement between the pretest and posttest, standardized score
improvement (Wilcox's $Q$), and time spent on the learning activity for each
experimental group.}
\label{tab:preposttests}
\vskip -1em
\end{table*}
A Shapiro-Wilk
test also showed that the posttest scores were non-normal
($p<.001$ for all three experimental groups), so we use a paired Mann Whitney U
test to test for learning gains within each group and found that all three
groups performed significantly better on the posttest than on the pretest ($p <
.001$ for all three groups). All three groups improved by between 9 and 10
rubric points on average, or between 32\% and 35\%.
%See Table
%\ref{tab:preposttests} for full details and Figure~\ref{fig:all-violin} to see
%plots of the score distributions.
Because of
concerns over the
use of Cohen's $d$ for effect sizes between non-normal
distributions~\cite{wilcox2019robust}, we use
Wilcox's $Q$ as our standardized effect size for learning gains and find that
students improved by a standardized effect size between $0.66$ and $0.76$.
We also performed a Kruskal-Wallis test on the posttest scores between groups,
and we do not reject the null hypothesis that students from each experimental
group performed equally well on the posttest ($\chi^2=0.54$, $p=0.76$).

Thus, we conclude that students in all three experimental groups improved at
proof by induction by reading the lecture notes and completing their
respective learning activities and that no one group learned significantly
more than the others.

\subsection{Time Spent on Learning Activity}
\begin{figure}
	\includegraphics[width=\columnwidth]{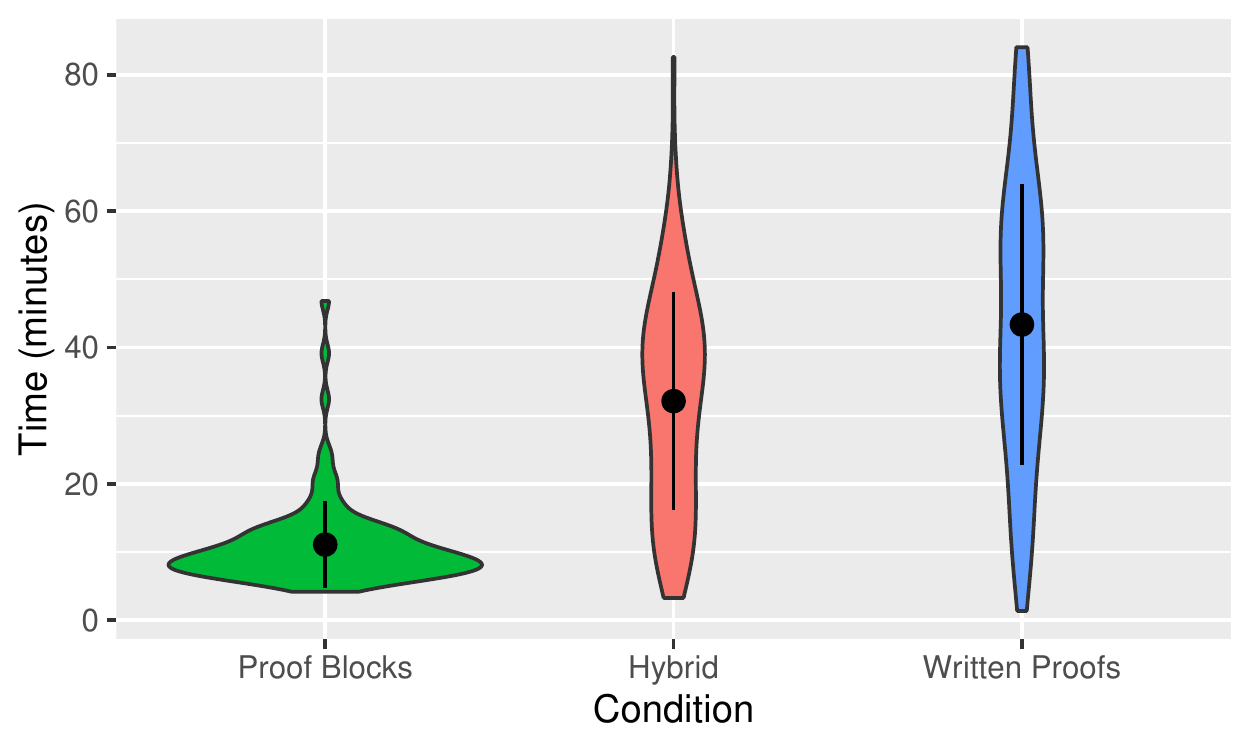}
	\caption{Distribution of time spent on each learning activity by each of
	the experimental groups. Students who completed the \tool{} activity were
	able to complete their activity much more quickly.}
	\label{fig:activity-times}
	\vskip -1em
\end{figure}

Next, to answer \textbf{RQ2}, we examine the amount of time that students from
each group spent on the learning activity (see
Figure~\ref{fig:activity-times}). Pairwise $t$-tests show that students in the
Written
Proof activity took significantly longer on their learning activity that
students in the Hybrid and Proof Blocks conditions, and that students in the
Hybrid condition took significantly longer than students in the Proof Blocks
condition ($p < 0.001$ in all cases).
Students in the Proof Blocks condition completed their activity about four
times faster on average than students in the Written Proofs condition. Full
details of the time differences and statistical tests can be seen in Table
\ref{tab:timestats}.
Using this together with the above result that students in all conditions
learned equally, we conclude that students who completed the \tool{} and Hybrid
activities learned more per unit time than students who completed the
proof-writing activity.

\begin{table}
\begin{tabular}{c|cc}
& Mean Time
&  \\
& Difference in minutes
& Cohen's $d$ \\ \hline
Hybrid $-$ Proof Blocks & 21.04 & 1.71 \\
Written $-$ Proof Blocks & 32.27 & 2.09 \\
Written $-$  Hybrid & 11.23 & 0.61 \\
\end{tabular}
\caption{Difference in time spent on the learning activity between each pair of
experimental groups. The students who completed the written proofs learning
activity took the longest by far, but did not learn any more than the other
groups.}
\label{tab:timestats}
\vskip -1em
\end{table}

\begin{figure}
\includegraphics[width=\columnwidth]{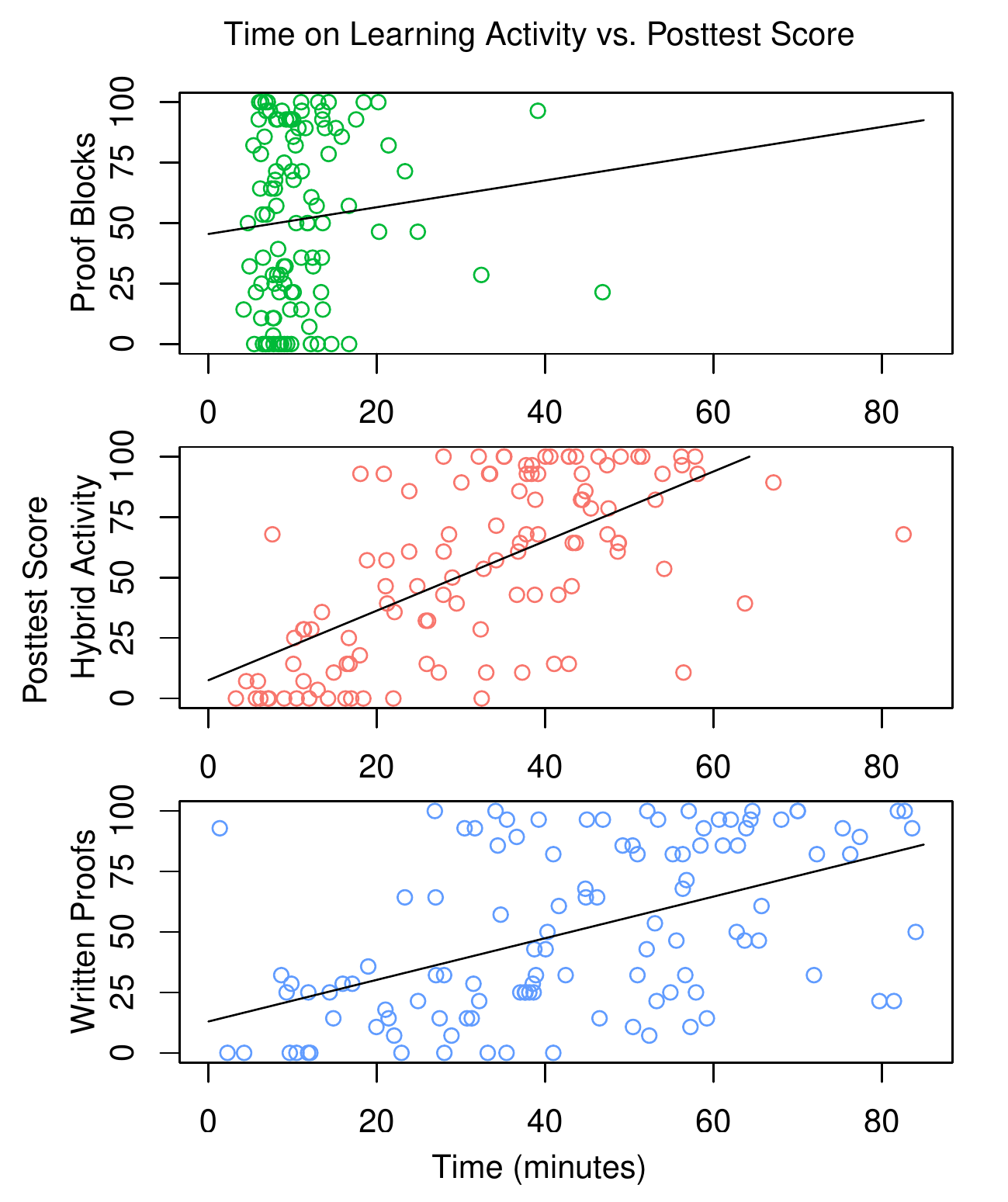}
\caption{For the \tool{} learning activity, time on task and posttest scores
were uncorrelated, and for the hybrid and written proof activities, they were
moderately correlated.}
\label{fig:time-vs-posttest}
	\vskip -1em
\end{figure}

Within a given learning condition, there is a relatively weak relationship
between time spent on task and posttest score. For the \tool{} learning
activity, time spent on the
activity and posttest score were essentially uncorrelated ($p=0.32$, $r=0.10$,
95\% CI [$-0.09, 0.28$]). The time spent on the activity and posttest scores
were moderately correlated for the hybrid activity ($p<0.001$, $r=0.64$, 95\%
CI [$0.52, 0.74$]) and for the written proofs acivity ($p<0.001$, $r=0.51$, 95\%
CI [$0.36, 0.63$]). These correlations are visualized in
Figure~\ref{fig:time-vs-posttest}.

Many students took much longer than expected on the `Written Proofs' learning
activity, resulting in some students possibly not having time to complete the
practice test. As a robustness check, we re-ran
the analysis, dropping all students who may not have had time to finish the
practice test, which was all those who took longer than 55 minutes or 0.56
standard deviations above the mean. This was true for 37 out of 113
students in the Written proofs condition. To obtain a fair comparison, we
likewise dropped all students who
were 0.56 standard deviations above mean time on the hybrid activity, which was
24 of the 112 students. We found all
results of the analysis to be the same. Since our robustness check shows the
results for the subset of the data would be identical to that of the entire
data set, for simplicity we present the results over the whole data set.

\subsection{Conditional Analyses}
As a further robustness check, we compared only students who responded
similarly on the familiarity survey. We re-ran all analyses, considering only
students who had never heard of proof by induction ($n=96$), and considering
only students who were somewhat familiar with proof by induction ($n=214$), and
found the same results in both cases. We did not re-run the analysis for
students who were very familiar with proof by induction, because there are too
few of them to have meaningful results ($n=22$).

We were also interested if students from the different experimental groups
learned different parts of the proof by induction at different amounts. To test
for this, we re-ran all analysis on each section of the rubric individually:
the base case, the inductive hypothesis, and inductive step. All analyses
showed exactly the same results, indicating that the learning happened roughly
equally across all parts of the rubric, across all three experimental groups.
Rerunning all analyses on each of the two test questions separately also gave
all the same results, showing that the learning was not different across the
two test questions.

\section{Discussion and Limitations}
Students who read lecture notes completed the \tool{} and
hybrid activities learned as much as students who read lecture notes and completed
the Written Proofs
activity, but in a shorter amount of time.
This result gives a foundation for future work about student learning using
\tool.
%We also saw that students who were given \tool{} problems had an easier time
%getting started on their learning activity:

%For example, how do learning gains compare for students
%using \tool{} problems with and without distractor lines? When is the ideal
%time to transition from practicing with \tool{} to practicing with written
%proofs? What is the ideal way to have \tool{} problems situated within course
%content? Are \tool{} also useful for helping students learn to write proofs on
%more advanced topics such as in a computational theory or algorithms course?

One limitation of our study is that it has limited ecological
validity. For example, it would be useful to explicitly study ways to situate
\tool{} problems within course content, as Weinmann et al. did with
faded Parsons problems~\cite{weinman2021improving}. Another limitation is that
because the learning of all three
experimental groups was the same, there is possibility that the student
learning happened entirely from reading the lecture notes, and not from the
learning activities. We will address this
concern in a follow-up study in which one of the experimental groups completes
either no learning activity, or a learning activity on an unrelated topic.
%\todo{tweak this next sentence to sound better}.
We are also not able to comment on learning saturation---how many \tool{} or
written proof problems must a student complete before they aren't learning any
more from each additional problem? And what is the most effective way to help
students continue to improve after they have completed a few \tool{} problems?
Due to the conditions of our IRB protocol, we do not have any demographic
information about our study participants.

\section{Conclusions}
In this work, we measured student learning
gains across different learning activities in a randomized controlled trial.
Our experiment showed that students in the early phases of learning a new type
of proof learned just as much reading lecture notes and using \tool{} as reading
lecture notes and writing proofs on their own,
but in far less time. Future work should continue to investigate the merits of
\tool{} as a learning tool in various contexts and how it can be extended to
further scaffold and improve learning.

%There is also the matter of testing learning gains on different types of
%content.
%Would it be a more useful tool earlier on in the semester, when students are
%learning what a proof is for the very first time?
%Could it help students in a Calculus class learn about writing their
%justifications
%for their solutions to optimization problems or $\delta$-$\epsilon$ proofs?

\begin{acks}
We would like to give a huge thanks to Dave Mussulman and the rest
of the staff at the computer based testing facility for helping us use their
facility to run our experiment. Seth Poulsen was supported by an NSF Graduate
Research Fellowship.
\end{acks}

\newpage
\balance
\bibliographystyle{ACM-Reference-Format}
\bibliography{references}

\end{document}